\documentclass{PoS}

\def\beq{\begin{equation}}
\def\eeq{\end{equation}}
\def\bsp#1\esp{\begin{split}#1\end{split}}
\def\beqa{\begin{eqnarray}}
\def\eeqa{\end{eqnarray}}

\def\sect#1{Sec.~{\ref{#1}}}

\def\eqn#1{eq.~(\ref{#1})}

\def\eqns#1#2{eqs.~(\ref{#1}) and~(\ref{#2})}

\def\ord{{\cal O} }
\def\eps{\epsilon}
\def\e{\epsilon}

\title{Infrared singularities and the high-energy limit}

\ShortTitle{The infrared structure of gauge amplitudes in the high-energy limit}

\author{Vittorio Del Duca\\
            INFN, Laboratori Nazionali di Frascati\\
            E-mail: \email{delduca@lnf.infn.it}}

\author{Claude Duhr\\
            ETH, Zurich\\
            E-mail: \email{duhrc@itp.phys.ethz.ch}}

\author{Einan Gardi\\
            Tait Institute, University of Edinburgh\\
            E-mail: \email{Einan.Gardi@ed.ac.uk}}

\author{\speaker{Lorenzo Magnea}\\
        University of Torino and INFN, Sezione di Torino\\
        E-mail: \email{magnea@to.infn.it}}

\author{Chris D. White\\
            University of Glasgow\\
            E-mail: \email{Christopher.White@glasgow.ac.uk}}

\abstract{We review recent results on the high-energy limit 
of gauge amplitudes, which can be derived from the universal 
properties of their infrared singularities. Using the dipole formula, a 
compact ansatz for infrared singularities of massless gauge 
amplitudes, and taking the high-energy limit, we provide a
simple expression for the soft factor of a generic high-energy
amplitude, valid to leading power in $t/s$ and to all logarithmic
orders. This gives a direct and general proof of 
leading-logarithmic Reggeization for infrared divergent 
contributions to the amplitude, and it shows how Reggeization
breaks down at NNLL level. We further show how the dipole 
formula constrains the high-energy limit of multi-particle 
amplitudes in multi-Regge kinematics, and how, on the other 
hand, Regge theory constrains possible corrections 
to the dipole formula.}

\FullConference{10th International Symposium on Radiative Corrections 
(Applications of Quantum Field Theory to Phenomenology)\\
                 September 26-30, 2011\\
                 Mamallapuram, India}

\begin{document}

\section{Introduction}
\label{INT}

The high-energy limit of gauge theory scattering amplitudes and cross sections has
been a source of interest for theory and phenomenology for more than half a century,
starting with studies performed in the context of Regge theory, long before the 
construction of the Standard Model of particle physics~\cite{ELP}. In the context of 
Perturbative QCD, and using contemporary language, the problem may be formulated 
as follows. Consider the key example of a four-point scattering amplitude, characterized 
by the Mandelstam invariants $s$, $t$ and $u$, and by particle masses $m_i$, satisfying
\beq
s + t + u = \sum_{i = 1}^4 m_i^2 \, .
\label{Mandel}
\eeq
The high-energy limit is the limit in which the squared center-of-mass energy $s$ 
is much larger than other kinematic invariants, compatibly with the constraint in 
\eqn{Mandel}. In the following, we concentrate exclusively on massless scattering,
$m_i^2=0$, and one is then led to consider 
(in the center-of-mass frame) either forward scattering ($s \gg - t$, $u \sim - s$), 
or backward scattering ($s \gg - u$, $t \sim - s$).  As expected in general for 
renormalizable field theories, in the presence of two disparate scales, say $s$ 
and $|t|$, amplitudes are dominated order by order by logarithmic contributions, 
which eventually spoil the (asymptotic) convergence of perturbation theory and 
need to be resummed. Regge theory suggests, under suitable assumptions, that 
this resummation can be achieved by replacing the $t$-channel propagator of the 
particle responsible for the leading power contribution to the high-energy limit 
at Born level with its `Reggeized' counterpart, according to 
\beq
  \frac{1}{t} \, \longrightarrow \, \frac{1}{t} \, 
  \left( \frac{s}{- t} \right)^{\alpha(t)} \, ,
\label{propregg}
\eeq
where $\alpha(t)$ is the Regge trajectory for the chosen $t$-channel exchange.
In Perturbative QCD, the Regge ansatz, \eqn{propregg}, can be studied explicitly,
and `Reggeization' can be proved, at least for selected amplitudes and to a given
logarithmic accuracy~\cite{BFL,Bogdan:2006af,Fadin:2006bj}. Consider for example 
gluon-gluon scattering. To leading logarithmic (LL) accuracy~\cite{BFL}, and to 
next-to-leading logarithmic (NLL) accuracy for the real part only~\cite{Fadin:2006bj}, 
one can show that the scattering amplitude obeys a Regge factorization of the form
\beq
  {\cal M}^{g g \rightarrow g g}_{a_1a_2 a_3 a_4} (s,t) 
  \, = \, 2 \, g_s^2 \, \frac{s}{t} \,
  \bigg[ (T^b)_{a_1a_3} C_{\lambda_1\lambda_3}(k_1, k_3) 
  \bigg] \, \left( \frac{s}{- t} \right)^{\alpha(t)} \,
  \bigg[ (T_b)_{a_2 a_4} C_{\lambda_2\lambda_4}(k_2, k_4) 
  \bigg] \,,
\label{Mgg}
\eeq
where we take $s = (k_1 + k_2)^2$ and $t = (k_1 - k_3)^2$, $T^a$ are color 
generators in the adjoint representation, $g_s$ is the strong coupling, and the 
functions $C_{\lambda_i \lambda_j} (k_i, k_j)$ are universal `impact factors' 
associated with gluons carrying momenta $k_i$ and helicities $\lambda_i$. 
Eq.~(\ref{Mgg}) represents a universal factorization since impact factors
depend only on the identity of the energetic partons interacting via $t$-channel
exchange, while the $s$ dependence is confined to the Reggeized propagator,
which in turn depends only on the identity of the particle being exchanged
in the $t$ channel. Once this factorization is established, one may compute
the various factors, and in particular the Regge trajectory $\alpha(t)$, using
perturbation theory. A crucial fact is that the Regge trajectory is IR 
divergent order by order, since it requires integrations over virtual corrections
which involve soft gluon exchanges. Using dimensional continuation (to $d
\equiv 4 - 2 \epsilon > 4$) to regulate the infrared, it is practical to expand 
the Regge trajectory in powers of the $d$-dimensional running 
coupling~\cite{Magnea:1990zb}, defined by the RG 
equation
\beq
  \mu \frac{\partial \alpha_s}{\partial \mu} \, = 
  \beta (\epsilon, \alpha_s) \, = \, - \, 2 \epsilon \alpha_s - 
  \frac{\alpha_s^2}{2 \pi} \, \sum_{n = 0}^\infty
  b_n \left( \frac{\alpha_s}{\pi} \right)^n \, ,
\label{running_coupling_RGE}
\eeq
where $\epsilon < 0$ and $b_0 = (11 C_A - 2 n_f)/3$. One then writes
\beq
  \alpha(t) \, = \, \frac{\alpha_s (- t, \epsilon)}{4 \pi} 
  \, \, \alpha^{(1)} +
  \left( \frac{\alpha_s (- t, \epsilon)}{4 \pi} \right)^2 
  \, \alpha^{(2)} + \, \ord \left( \alpha_s^3 \right) \, ,
\label{alphb}
\eeq
where the coupling is naturally evaluated at the scale $\mu = - t$, 
\beq
  \alpha_s (- t, \epsilon) = \left( \frac{\mu^2}{- t} 
  \right)^\epsilon \, \alpha_s (\mu^2) 
  + \, \ord \left( \alpha_s^2 \right) \, .
\label{rescal}
\eeq 
The first two perturbative coefficients of the gluon Regge trajectory are known
(see for example~\cite{Kuraev:1977fs,Fadin:1996tb,DelDuca:2001gu,
Sotiropoulos:1993rd,Korchemskaya:1996je}), and in the $\overline{\rm MS}$ 
scheme they are  given by
\beq
  \alpha^{(1)} \, = \, C_A \, 
  \frac{\widehat{\gamma}_K^{(1)}}{\epsilon} \, ,
  \qquad
  \alpha^{(2)} \, = \, C_A \left[ - {b_0 \over \eps^2} +
  \widehat{\gamma}_K^{(2)} \,
  {2 \over \eps} + C_A \left( {404 \over 27} - 2 \zeta_3 \right) 
  + n_f \left(- {56 \over 27} \right) \right] \, .
\label{eq:2loop}
\eeq
In \eqn{eq:2loop} we introduced the cusp anomalous 
dimension~\cite{Korchemsky:1987wg}, $\gamma_K^{(i)} (\alpha_s)$, 
where $i$ denotes the color representation (the adjoint representation 
in the present case), and we extracted the corresponding quadratic 
Casimir eigenvalue, writing $\gamma_K^{(i)} (\alpha_s) \equiv C_i \, 
\widehat{\gamma}_K (\alpha_s)$. The first coefficients of the universal 
function $\widehat{\gamma}_K (\alpha_s)$, which appear in 
\eqn{eq:2loop}, are well known and are given by
\beq 
  \widehat{\gamma}_K^{(1)} \, = \, 2 \, , \qquad 
  \widehat{\gamma}_K^{(2)} = \left({67 \over 18} -
  {\pi^2 \over 6} \right) C_A - {5\over 9} n_f \, .
\label{eq:beta}
\eeq
The fact that the Regge trajectory is IR divergent suggests that studies 
of the high-energy limit could be pursued by using the detailed knowledge of 
IR and collinear divergences that has been developed over the 
last decades. Notably, considerable progress was achieved in the last few years, 
which in particular led to an all-order ansatz~\cite{Gardi:2009qi,Becher:2009cu} 
for the anomalous dimension governing the exponentiation of IR and collinear 
poles in generic multi-particle massless scattering amplitudes. In the following, we 
will first describe this ansatz (the `dipole formula'), briefly discussing its properties 
and limitations, and then we will show how it can be used to improve our 
understanding of the high-energy limit~\cite{DelDuca:2011xm}. We will show 
that, at leading power in $|t|/s$ and to arbitrary logarithmic accuracy, IR divergences 
are generated by a soft operator with a simple color and kinematic structure, and 
we will use this result to prove Reggeization of IR divergent contributions to the 
amplitude, at LL level, for the scattering of particles belonging to generic color 
representations. Our soft operator, however, is accurate to all logarithmic orders, 
so we can go beyond LL: we find that standard Regge factorization breaks down 
at NNLL, and we give a general expression for the leading Reggeization-breaking 
operator. We then show how our method applies to multi-particle scattering in 
Multi-Regge kinematics, and we point out how the Regge limit constrains 
potential corrections to the dipole ansatz.

\section{Dipoles}
\label{DIP}

Our understanding of IR and collinear divergences in gauge theory 
amplitudes rests on a powerful factorization theorem, which is the final result 
of many years of theoretical work~\cite{Sterman:1995fz}. In order to state the 
theorem, let us first note that a scattering amplitude in a non-abelian gauge 
theory is a vector in the space of the available color configurations, which is a 
subspace of the tensor product of the color representations of the $n$ particles 
participating in the scattering process. Choosing a basis of color tensors $c_J$ 
in this space we may write 
\beq
  {\cal M}_{a_1\ldots a_L} \left(\frac{p_i}{\mu}, \alpha_s (\mu^2),
  \epsilon \right) \, = \,
  \sum_J  {\cal M}_J \left( \frac{p_i}{\mu}, \alpha_s (\mu^2), 
  \epsilon \right) \, (c_J)_{a_1 \ldots a_L} \, ,
\label{ampcol}
\eeq
In the following we will consider massless fixed-angle scattering amplitudes,
so that $p_i^2 = 0$, and all invariants $p_i \cdot p_j \gg \Lambda_{\rm QCD}$ 
are taken to be of the same parametric size. The factorization theorem then 
states~\cite{Dixon:2008gr} that each component of the vector ${\cal M}$ can 
be written as a product
\beqa
  {\cal M}_{J} \left(\frac{p_i}{\mu}, \alpha_s (\mu^2), 
  \epsilon \right)  & = & \sum_K {\cal S}_{J K} \left(\beta_i  
  \cdot \beta_j, \alpha_s (\mu^2), 
  \epsilon \right) \,  H_{K} \left( \frac{2 p_i \cdot p_j}{\mu^2},
  \frac{(2 p_i \cdot n_i)^2}{n_i^2 \mu^2}, \alpha_s (\mu^2),
  \epsilon \right) \nonumber \\ & & \times \,
  \prod_{i = 1}^L \, \frac{{\displaystyle J_i 
  \left(\frac{(2 p_i \cdot n_i)^2}{n_i^2 \mu^2},
  \alpha_s (\mu^2), \epsilon \right)}}{{\displaystyle {\cal J}_i 
  \left(\frac{2 (\beta_i \cdot n_i)^2}{n_i^2}, \alpha_s (\mu^2), 
  \epsilon \right)} \,} \,\, .
\label{fact_massless}
\eeqa
Here we have introduced the light-like four velocities $\beta_i^\mu \equiv
p_i^\mu/Q$, with $Q \sim \mu$ an arbitrary scale, and the auxiliary four-vectors
$n_i^\mu$, $n_i^2 \neq 0$, to be discussed below. The matrix ${\cal S}$ is
responsible for soft divergences, and it acts as a color operator on the vector
of finite functions $H$. It is defined by the expectation value of a product of
Wilson lines, as
\beq
  \left( c^J \right)_{\{a_k\}} {\cal S}_{J K} \left(\beta_i 
  \cdot \beta_j, \alpha_s (\mu^2), \epsilon \right) 
  \equiv \sum_{\{b_k\}} \, \, \left\langle 0 \left| 
  \, \prod_{k = 1}^L \Big[ \Phi_{\beta_k} (\infty, 0) \Big]_{a_k b_k} 
  \, \right| 0 \right\rangle_{ \!\! {\rm ren.}}  \, 
  \left( c^K \right)_{\{b_k\}} \, ,
\label{softcorr_massless}
\eeq
where
\beq
  \left[ \Phi_{\beta_l} (\infty, 0) \right]_{a_l b_l} \, \equiv \, 
  \left[ {\cal P} \exp \left( {\mathrm i} 
  g_s \int_0^{\infty} d t \, \beta_l \cdot {A} (t \beta_l ) \right) 
  \right]_{a_l b_l} \, .
\label{Wilson_line_def}
\eeq
Collinear divergences associated with each external hard particle are collected
in the jet functions $J_i$, which are defined in a gauge-invariant way with the help
of the auxiliary vectors $n_i$. As an example, for quark jets one writes
\beq
  \overline{u}(p_l) \, J_l \left( \frac{(2 p_l 
  \cdot n_l)^2}{n_l^2 \mu^2}, 
  \alpha_s(\mu^2), \e \right) \, = \, \langle p_l \, | \, 
  \overline{\psi} (0) \, \Phi_{n_l} (0, - \infty) \,  | 0 \rangle\, .
\label{Jdef}
\eeq
In taking the product of soft and collinear factors, one double-counts the soft-collinear
region. The problem is easily fixed by dividing each particle jet by its own eikonal
approximation${\cal J}_i$, which is obtained simply by replacing the field $\psi$ in
\eqn{Jdef} with a Wilson line in the direction of the corresponding momentum $p_i$. 
Note that the dependence on the `factorization vectors' $n_i$ must cancel between 
the jet factors and the hard functions. Upon modifying the hard function in order to 
effect this cancellation, the factorization theorem in \eqn{fact_massless} can be 
summarized~\cite{Becher:2009cu} in a more compact and elegant form as
\beq
  {\cal M} \left(\frac{p_i}{\mu}, \alpha_s (\mu^2),  \epsilon 
  \right) \, = \, Z \left(\frac{p_i}{\mu_f}, \alpha_s(\mu_f^2),  
  \epsilon \right) \, \, {\cal H} \left(\frac{p_i}{\mu}, 
  \frac{\mu_f}{\mu}, \alpha_s(\mu^2), \epsilon \right) \, ,
\label{Mfac}
\eeq
where we introduced a factorization scale $\mu_f$, and where $Z$ acts as a color
operator on the vector of modified hard functions ${\cal H}$. The operator $Z$, 
which is built out of fields and Wilson lines, is multiplicatively renormalizable, and 
thus obeys a (matrix) RG equation of the form
\beq
  \mu \frac{d}{d \mu} \, Z \left(\frac{p_i}{\mu}, \alpha_s(\mu^2), \epsilon \right) \, = \, 
  - \,  Z \left(\frac{p_i}{\mu}, \alpha_s(\mu^2), \epsilon \right) \,
  \Gamma \left(\frac{p_i}{\mu}, \alpha_s(\mu^2) \right) \, ,
\label{RG}
\eeq
with a finite anomalous dimension matrix $\Gamma$. In dimensional regularization,
\eqn{RG} can be easily solved to yield
\beq
  Z \left(\frac{p_l}{\mu}, \alpha_s(\mu^2), \epsilon \right) \, = \,  
  {\cal P} \exp \left[ \frac{1}{2} \int_0^{\mu^2} \frac{d \lambda^2}{\lambda^2} \, \,
  \Gamma \left(\frac{p_i}{\lambda}, \alpha_s(\lambda^2) \right) \right] \, .
\label{RGsol}
\eeq
Clearly, our understanding of IR and collinear divergences to all 
orders hinges upon the knowledge of the anomalous dimension matrix 
$\Gamma$. The dipole formula is an ansatz for this matrix, which represents 
the simplest solution to a set of exact equations satisfied by $\Gamma$ 
as a consequence of factorization and of conformal invariance of light-like 
Wilson lines~\cite{Gardi:2009qi,Becher:2009cu}. It reads
\beq
  \Gamma_{\rm dip}  \left(\frac{p_i}{\lambda}, \alpha_s(\lambda^2) \right) \, = \,
  \frac{1}{4} \, \widehat{\gamma}_K \left(\alpha_s (\lambda^2) \right) \,
  \sum_{(i,j)} \ln \left(\frac{- s_{ij}}{\lambda^2} 
  \right) {\bf T}_i \cdot {\bf T}_j \, - \, \sum_{i = 1}^L
  \gamma_{J_i} \left(\alpha_s (\lambda^2) \right) \, ,
\label{sumodipoles}
\eeq
where $- s_{ij} = 2 |p_i \cdot p_j| {\rm e}^{- {\rm i} \pi \lambda_{i j}}$, with 
$\lambda_{i j} = 1$ if particles $i$ and $j$ both belong to either the initial or 
the final state, and $\lambda_{i j} = 0$ otherwise. The color structure 
is expressed in a basis-independent way by the color generators ${\bf T}_i$,
which act on hard parton $i$ as gluon insertion operators, and $\gamma_{J_i}$
are color singlet anomalous dimensions for the jets. The crucial feature of 
\eqn{sumodipoles} is that it involves only pairwise correlations between 
hard partons, an extremely drastic simplification with respect to the expected
level of complexity. A side effect of this simplification is the fact that the matrix 
structure of $\Gamma_{\rm dip}$ is fixed at one loop, and radiative corrections
enter only through the anomalous dimensions $\widehat{\gamma}_K$ and 
$\gamma_{J_i}$: as a consequence, the path ordering in \eqn{RGsol} becomes
immaterial. The dipole formula reproduces all known results for IR divergences 
of massless gauge theory amplitudes, and in principle it could be the 
definitive solution of the problem. We know however that \eqn{sumodipoles}
may receive, at sufficiently high orders, precisely two classes of 
corrections~\cite{Gardi:2009qi,Becher:2009cu}. First of all, \eqn{sumodipoles} 
was derived assuming that the cusp anomalous dimensions in 
representation $i$ remains proportional to the quadratic Casimir 
eigenvalue $C_i$ to all orders in perturbation theory. This proportionality
("Casimir scaling") is established through explicit calculations only up to three 
loops, and in principle corrections proportional to higher-order Casimir operators
might arise starting at four loops (although arguments were given~\cite{Becher:2009cu} 
that at four loops this does not actually happen).  The second class of possible 
corrections to \eqn{sumodipoles} arises because the conformal invariance of light-like 
Wilson lines cannot constrain the dependence of $\Gamma$ on conformally invariant 
cross-ratios of momenta such as $\rho_{ijkl} \equiv (p_i \cdot p_j p_k \cdot p_l)/(p_i 
\cdot p_k p_j \cdot p_l)$. If such corrections are present, they modify the anomalous 
dimension matrix according to
\beq
  \Gamma \left(\frac{p_i}{\lambda}, \alpha_s(\lambda^2) \right) \, = \,
  \Gamma_{\rm dip} \left(\frac{p_i}{\lambda}, \alpha_s(\lambda^2) \right) 
  \, + \, \Delta \left( \rho_{ijkl}, \alpha_s (\lambda^2) \right) \, .
\label{Delta}
\eeq
The `quadrupole' correction $\Delta$ can be non-vanishing starting at three
loops and when at least four hard particles are present. It has been shown to be
tightly constrained not only by factorization, but also by collinear limits, Bose
symmetry and transcendentality requirements. A few examples of viable 
functions at three loops have however been constructed~\cite{Dixon:2009ur}, 
and the general structure of possible corrections at four loops has been 
studied~\cite{Vernazza:2011aa}. As we will see shortly, the high-energy limit 
provides further nontrivial constraints, which come tantalizingly close to showing 
that $\Delta$ should vanish, at least at three loops.

\section{High-energy}
\label{HE}

Since the dipole formula is an ansatz applicable  to all orders and for any number 
of hard particles, it is natural to study its high-energy limit. Note that, strictly 
speaking, the fixed-angle assumption under which the dipole formula was derived 
breaks down as $|t|/s \to 0$. The ensuing corrections however must take the form
of logarithms of $s/(-t)$, with coefficients that remain finite as $\epsilon \to 0$, since 
the high-energy limit cannot generate new divergent contributions. In order to write 
the result in a compact and appealing form, it is useful to introduce color operators 
associated with each Mandelstam channel~\cite{Dokshitzer:2005ig}. Explicitly
\beq
  {\bf T}_s \, = \, {\bf T}_1 + {\bf T}_2 \, = \, 
  - \left( {\bf T}_3 + {\bf T}_4 \right) \, , \qquad
  {\bf T}_t \, = \, {\bf T}_1 + {\bf T}_3 \, = \, 
  - \left( {\bf T}_2 + {\bf T}_4 \right) \, 
\label{Tstudef}
\eeq
and similarly for ${\bf T}_u$. In \eqn{Tstudef}, we enforced color conservation, 
expressed in this context by $\sum_i {\bf T}_i = 0$. For scattering between 
particles in generic color representations, the color operators in \eqn{Tstudef} 
satisfy a constraint reminiscent of \eqn{Mandel},
\beq
  {\bf T}_s^2 + {\bf T}_t^2 + {\bf T}_u^2 \, = \, 
  \sum_{i = 1}^4 C_i \, .
\label{colcon2}
\eeq
We consider now the limit $|t| \ll s$, which implies $u \sim - s$. The energy 
dependence of the four-point amplitude simplifies drastically in this limit, and 
in particular the IR operator $Z$ factorizes, to leading {\it power} in
$|t|/s$, as
\beq
  Z \left( \frac{p_i}{\mu}, \alpha_s (\mu^2), \epsilon \right)
  \, = \, \widetilde{Z} \left( \frac{s}{t}, \alpha_s (\mu^2), \epsilon 
  \right) \, Z_{\bf 1} \left( \frac{t}{\mu^2}, \alpha_s (\mu^2), 
  \epsilon \right) \, .
\label{Zfac}
\eeq
The crucial feature of \eqn{Zfac} is that the factor $Z_{\bf 1}$, which is independent
of $s$, is proportional to the unit matrix in color space, while the non-trivial
color dependence, contained in the factor $\widetilde{Z}$, is remarkably simple.
One finds
\beq
  \widetilde{Z} \left( \frac{s}{t}, \alpha_s (\mu^2), \epsilon \right) 
  \, = \, \exp \left\{ K \Big(\alpha_s (\mu^2), \epsilon \Big)
  \Bigg[ \ln \left( \frac{s}{- t} \right) {\bf T}_t^2 + 
  {\rm i} \pi \, {\bf T}_s^2 \Bigg] \right\} \, ,
  \label{Ztildedef}
\eeq
where the function $K(\alpha_s, \epsilon)$ is a simple integral of the cusp
anomalous dimension, which plays a role in several different QCD applications,
given by
\beq
K \Big(\alpha_s (\mu^2), \epsilon \Big) \, \equiv \,  
  - \frac14 \int_0^{\mu^2} \frac{d \lambda^2}{\lambda^2} \, 
  \widehat{\gamma}_K \left(\alpha_s(\lambda^2, \epsilon) 
  \right) \, . 
  \label{Kdef} 
\eeq
We note that the special role played by the cusp anomalous dimension in the 
high-energy limit was previously demonstrated in~\cite{Sotiropoulos:1993rd,
Korchemskaya:1996je}; in particular, Ref.~\cite{Korchemskaya:1996je} derived
an expression analogous to \eqn{Kdef}, and used it to compute the gluon Regge 
trajectory at NLL. One readily sees, in fact, that the simple dependence of 
\eqn{Ztildedef} on the high-energy logarithm $\ln(s/(-t))$ has immediate consequences 
on Reggeization, at least for divergent contributions to the amplitude. At LL level, 
one can obviously discard the imaginary part of the exponent of \eqn{Ztildedef}, 
and the matrix element in \eqn{Mfac} takes the form
\beq
  {\cal M} \left( \frac{p_i}{\mu}, 
  \alpha_s(\mu^2), \epsilon \right) \, = \, 
  \exp \left\{ K \Big( \alpha_s (\mu^2), 
  \epsilon \Big) \, \ln \left(\frac{s}{- t} \right) \,
  {\bf T}_t^2 \right\} \, Z_{\bf 1} \, {\cal H} \left( \frac{p_i}{\mu}, 
  \alpha_s(\mu^2), \epsilon \right) \, ,
\label{Mggdef}
\eeq
This automatically implies (LL) Reggeization, for any representation content,
under the sole assumption that the amplitude be dominated at lowest order
and at leading power in $|t|/s$ by the exchange of a specific color state in the
$t$ channel. If this is the case the color operator ${\bf T}_t^2$ can be replaced
by its Casimir eigenvalue $C_t$, and the color state exchanged in the $t$ 
channel is found to Reggeize, with a Regge trajectory given by the function
$C_t K(\alpha_s, \epsilon)$. For example, for gluon-gluon scattering one finds
\beq
  {\bf T}_t^2 \, {\cal H}^{gg\rightarrow gg} \, =
  \, C_A \, {\cal H}^{gg\rightarrow gg}_t \, + \, \ord{\left(|t|/s \right)} \, ,
\label{eigen}
\eeq
which implies
\beq
  {\cal M}^{gg\rightarrow gg}  \, = \,
  \left(\frac{s}{-t}\right)^{C_A \, K \left(\alpha_s (\mu^2),
  \epsilon \right)} Z_{\bf 1} \, {\cal H}^{gg\rightarrow gg}_t \, .
\label{Mgg3}
\eeq
Computing the integral in \eqn{Kdef}, one easily recovers the divergent terms in 
\eqn{eq:2loop}. It is easy to see, however, that the same reasoning applies in full 
generality to different $t$-channel exchanges.

Since our master formula, \eqn{Ztildedef}, is valid to all logarithmic orders for 
divergent contributions, it is straightforward to go beyond LL accuracy. One 
may start by applying the Baker-Campbell-Hausdorff formula to write the operator 
$\widetilde{Z}$ as  a product of exponentials of decreasing logarithmic weight, as
\beqa
  && \hspace{-1mm} \widetilde{Z} \left( \frac{s}{t}, \alpha_s (\mu^2) , 
  \epsilon \right)  \, =  \, \left( \frac{s}{- t} \right)^{K \, \, {\bf T}_t^2} \, \, 
  \exp \left\{ {\rm i} \, \pi \, K \, \, {\bf T}_s^2 \right\} \,
  \exp \left\{ - \, {\rm i} \, \frac{\pi}{2}  \, K^2 \, 
  \ln \left(\frac{s}{- t} \right) \,  
  \Big[{\bf T}_t^2, {\bf T}_s^2 \Big] \right\} \label{Zcomm}
   \\ & & \hspace{-1mm} \times \, 
  \exp \left\{ \frac{K^3}{6} \left(- 2 \pi^2 \ln 
  \left(\frac{s}{- t}\right) 
  \Big[ {\bf T}_s^2, \big[{\bf T}_t^2, {\bf T}_s^2 \big] \Big] 
  \, + \,  {\rm i} \pi \, \ln^2 \left(\frac{s}{- t} \right) 
  \Big[{\bf T}_t^2, \big[{\bf T}_t^2, {\bf T}_s^2 
  \big] \Big] \right) \right\} \,\, \exp \left\{ {\cal O} \left( K^4 \right) \right\} \, ,
  \nonumber
\eeqa
where $K = K(\alpha_s, \e)$. From \eqn{Zcomm}, it is apparent that color exchanges
that are not diagonal in the $t$-channel basis become relevant at NLL accuracy for
the imaginary part of the amplitude (as exemplified by the presence of the color
operator ${\bf T}_s^2$ in the second factor on the {\it r.h.s.} of \eqn{Zcomm}), and 
for the real part of the amplitude at NNLL accuracy. This implies a breakdown of
Regge factorization, as given in \eqn{Mgg}. A general form of the leading 
Reggeization-breaking operator at NNLL for the real part of the amplitude is 
easily read off from \eqn{Zcomm} and is given by
\beq
  {\cal E} \left( \frac{s}{t}, \alpha_s, 
  \epsilon \right) \equiv - \, \frac{\pi^2}{3} \,
  {K^3 (\alpha_s, \epsilon)} \, \ln \left(\frac{s}{- t} \right) 
  \Big[{\bf T}_s^2, \big[{\bf T}_t^2, {\bf T}_s^2 \big] \Big] \, ,
\label{NNLL}
\eeq
where we note that the expansion of $K^3(\alpha_s, \e)$ starts out  at ${\cal O}
\left( \alpha_s^3/\e^3 \right)$.

The remarkable simplicity of \eqns{Zfac}{Ztildedef} suggests that IR divergences 
of multi-particle amplitudes might similarly simplify in a suitably defined high-energy 
limit. The relevant limit is well-known~\cite{DelDuca:1995hf}, and is characterized by 
strongly ordered rapidities $y_i$ of the emitted particles, while transverse momenta 
$k_i^\perp$ are of the same parametric size. This configuration is called `Multi-Regge' 
kinematical regime, characterized by
\beq
  y_3 \gg y_4 \gg \ldots \gg y_L \, \qquad 
  |k_i^\perp| \simeq |k_j^\perp| \quad \forall i,j  \, .
\label{rapord}
\eeq
In this limit, the relevant Mandelstam invariants can be written as
\beqa
  - \, s & \equiv & - \, s_{12} \, \simeq \, \, |k_3^\perp| 
  \, |k_L^\perp| \, e^{- {\rm i} \pi } \,
  {\rm e}^{y_3 - y_L} \, , \quad
  - \, s_{1 i} \, \simeq \,  \, |k_3^\perp| \, |k_i^\perp| 
  \, {\rm e}^{y_3 - y_i} \, , \quad
  - \, s_{2 i} \, \simeq \,  \, |k_L^\perp| \, |k_i^\perp| 
  \, {\rm e}^{y_i - y_L} \, , \nonumber \\ 
  - \, s_{i j} & \simeq &  \, |k_i^\perp| \, |k_j^\perp| 
  \, {\rm e}^{y_i - y_j} \, e^{- {\rm i} \pi } \,
  \quad  \quad \quad ( 3 \leq i < j \leq L) \, ,
\label{invars}
\eeqa
so that logarithms of ratios of kinematical invariants are dominated by rapidity 
differences. Taking the Multi-Regge limit in \eqn{sumodipoles} yields once again 
a factorized expression, in which the dominant kinematic dependence is 
characterized by $t$-channel color exchanges. One finds~\cite{DelDuca:2011xm}
\beq
  Z \left( \frac{p_l}{\mu}, \alpha_s (\mu^2), \epsilon \right)
  \, = \, \widetilde{Z}^{\rm MR} \Big( \Delta y_k, 
  \alpha_s (\mu^2), \epsilon \Big) \, \, Z_{\bf 1}^{\rm MR} 
  \left( \frac{|k_i^\perp|}{\mu}, 
  \alpha_s (\mu^2), \epsilon \right) \, ,
\label{ZMR} 
\eeq
where the transverse-momentum dependent factor $Z_{\bf 1}^{\rm MR}$ 
is again proportional to the unit matrix in color space. Non-trivial color
exchanges are encoded into the operator
\beq
  \widetilde{Z}^{\rm MR} \Big( \Delta y_k, \alpha_s (\mu^2), 
  \epsilon \Big) \, = \, \exp \left\{ K \Big( \alpha_s (\mu^2), 
  \epsilon \Big) \left[\sum_{k = 3}^{L-1} {\bf T}^2_{t_{k - 2}}
  \, \Delta y_k \, + \, {\rm i} \pi {\bf T}_s^2 \right] \right\} \, ,
  \label{ZtildeMR} 
\eeq
where we introduced the $t$-channel color matrices 
\beq
  {\bf T}_{t_k} \, = \, {\bf T}_1 + \sum_{p = 1}^k 
  {\bf T}_{p + 2} \, .
\label{Tkdef}
\eeq
It is important to note that the operators ${\bf T}_{t_k}^2$ commute with each 
other, so that it is always possible to choose a basis in color space in which they 
are all simultaneously diagonal. Given \eqn{ZtildeMR}, it is clear that the patterns 
of Reggeization, and Reggeization breaking, which have been discussed in the 
case of the four-point amplitude generalize to the Multi-Regge kinematics in a  
straightforward manner. Notice that it is not necessary that the tree-level hard 
amplitude be dominated by the exchange of a single color state across all the 
$t_k$ subchannels: if multiple representations contribute at leading power in the 
high-energy limit, they will separately Reggeize (at LL), with the expected Regge 
trajectories, as a consequence of \eqn{ZtildeMR}.

\section{Quadrupoles?}
\label{QUAD}

So far we have discussed the implications of the dipole formula on the 
high-energy limit of scattering amplitudes. It is important to note that known
results concerning the high-energy limit can, in turn,  be used to constrain
possible corrections to the dipole formula. As was outlined in \sect{DIP}, we 
already know that corrections to the dipole formula can only arise from violations 
of Casimir scaling for the cusp anomalous dimension, or in the form of functions
of the conformal cross ratios $\rho_{ijkl}$, denoted by $\Delta$ in \eqn{Delta}. 
In order to explore the existence of possible constraints on these functions arising
from the high-energy limit, let us focus on four-point amplitudes. In this case 
three different cross ratios can be constructed: $\rho_{1234}$, $\rho_{1342}$, 
$\rho_{1423}$, subject to the constraint $\rho_{1234} \rho_{1324} \rho_{1423} = 1$.
They are given by
\beq
  \rho_{1234}  \, =  \,  \left(\frac{s}{- t}\right)^2 \, {\rm e}^{- 2 {\rm i} \pi }
  \, ,  \qquad  
  \rho_{1342} \, = \, \left(\frac{- t}{s + t}\right)^2 \, ,  \qquad  
  \rho_{1423} \, = \, \left(\frac{s + t}{s}\right)^2 \, {\rm e}^{2{\rm i} \pi} \, , 
\label{L423_forward}
\eeq
where care was taken to retain the correct phases. In the high-energy limit, the logarithms 
of the cross ratios, $L_{ijkl} \equiv \log \rho_{ijkl}$ can be expressed, 
up to corrections suppressed by powers of $|t|/s \to 0$, in terms of 
the high-energy logarithm $L \equiv \log (s/(-t))$ as
\beq
  L_{1234}  \, = \, 2 ( L - {\rm i} \pi) \, , \qquad
  L_{1342}  \, = \, - 2 L \, , \qquad 
  L_{1423}  \, = \, 2 {\rm i} \pi \, .
\label{L423_forward_log}
\eeq
One may now examine the existing examples of functions $\Delta$ which
satisfy all previously examined constraints~\cite{Dixon:2009ur}. The simplest 
example, a symmetric polynomial in the logarithms $L_{ijkl}$, is given by
\beqa
\label{Delta_case212}
  \hspace{-5mm}
  \Delta^{(212)} (\rho_{ijkl}, \alpha_s) \, = \, 
  \left( \frac{\alpha_s}{\pi} \right)^3 \,
  {\bf T}_1^{a} {\bf T}_2^{b} {\bf T}_3^{c} {\bf T}_4^{d} \,
  \bigg[ && \! \! f^{ade} f^{cbe} \,
  L_{1234}^2 \, \Big(L_{1423} \, L_{1342}^{2} \, + \, 
  L_{1423}^{2} \, L_{1342} \Big) \\
  + && \! \! f^{cae} f^{dbe} \, 
  L_{1423}^2 \, \Big(L_{1234} \, L_{1342}^{2} \, + \, 
  L_{1234}^{2} \, L_{1342} \Big) \nonumber \\
  + && \! \! f^{bae} f^{cde} \,
  L_{1342}^{2} \, \Big(L_{1423} \, L_{1234}^{2} \, + \,
  L_{1423}^{2} \, L_{1234} \Big) \bigg] \, .
  \nonumber
\eeqa
In the high-energy limit (and using Jacobi identities for the
structure constants $f^{abc}$) one finds
\beqa
  \Delta^{(212)} (\rho_{ijkl}, \alpha_s) & = & 
  \left(\frac{\alpha_s}{\pi}\right)^3 \, {\bf T}_1^{a} {\bf T}_2^{b} 
  {\bf T}_3^{c} {\bf T}_4^{d} \, \,
  32 \, {\rm i} \, \pi \Big[ \Big( - L^4 - {\rm i} \pi L^3
  - \pi^2 L^2 - {\rm i} \pi^3 L \Big) f^{ade}f^{cbe}
  \nonumber \\ && \qquad + \, \,
  \Big(2 {\rm i} \pi L^3 - 3 \pi^2 L^2 - {\rm i} \pi^3 L \Big) 
  f^{cae}f^{dbe} \Big] \, + {\cal O} \left( |t/s| \right)\, .
\label{Delta_4_large_s}
\eeqa
Such a function cannot contribute to the anomalous dimension matrix $\Gamma$,
since it would generate super-leading high-energy logarithms, $\alpha_s^p L^q$, with
$q > p$, starting at three loops. One may easily show that all explicit examples 
constructed in Ref.~\cite{Dixon:2009ur} suffer from this problem: indeed, the problem 
appears to be rather generic, since the function $\Delta$ is constrained by collinear 
limits and Bose symmetry to contain a minimum number of factors behaving 
logarithmically at large $s$. In order to put together a viable example of a three-loop 
correction to the dipole formula, one would have to construct a linear combination of 
functions such as \eqn{Delta_case212}, each with the `wrong' high-energy behavior, 
arranging for the cancellation of the unwanted logarithms. Since gluon Reggeization 
is proven in general to NLL accuracy, in \eqn{Delta_4_large_s} one would need to 
cancel not only the super-leading $L^4$ contribution, but the leading ($L^3$) and 
next-to-leading ($L^2$) terms as well, since those are correctly predicted by the 
dipole formula, and would have a color structure incompatible with gluon Reggeization. 
So far, no explicit example of such a function has been constructed.

\section{Perspective}

We have summarized some recent developments concerning the IR structure
of scattering amplitudes in massless gauge theories, with applications to the widely
studied high-energy regime. The dipole formula, if correct, would be the first case 
in which the structure of a multi-particle non-abelian anomalous dimension is
understood exactly to all orders in perturbation theory: note that it applies to any
massless gauge theory and it is exact in the $1/N_c$ expansion. In the high-energy 
limit, the dipole formula implies Reggeization of IR-singular contributions to the 
amplitude at LL accuracy, and at NLL accuracy for the real part of the amplitude. 
Beyond NLL, it predicts the existence of specific Reggeization-breaking 
contributions, which can be explicitly computed in practical cases once the relevant 
hard sub-amplitudes are known. In turn, known results on the high-energy limit of 
scattering amplitudes constrain possible corrections to the dipole formula in a 
significant way, ruling out individually all viable examples which had been constructed 
so far, and it is conceivable that a definitive answer on the IR structure of multiparticle 
massless gauge theory amplitudes will become available in a not-too-far future.

\end{document}